\documentstyle[aas2pp4]{article}

\begin{document}

\title{Helical Strands in the Jet-like Narrow Line Region of ESO~428-G14}

\author{Heino Falcke, Andrew S. Wilson\altaffilmark{1}}
\affil{Astronomy Department, University of Maryland, College Park,
MD 20742-2421 \\(hfalcke@astro.umd.edu; wilson@astro.umd.edu)}
\altaffiltext{1}{Adjunct Astronomer, Space Telescope Science Institute}

\author{Chris Simpson}
\affil{Space Telescope Science Institute, 3700 San Martin Drive, Baltimore, MD 21218\\
(simpson@stsci.edu)}

\author{Gary A. Bower\altaffilmark{2}} 
\affil{Department of Physics \& Astronomy, Johns Hopkins University,
Baltimore, MD 21218\\(bower@pha.jhu.edu)}
\altaffiltext{2}{Current address: National Optical Astronomy
Observatories, P.O.~Box 26732, Tucson, AZ 85726}

\begin{abstract}
We present HST/WFPC2 images of the narrow line region (NLR) of the
Seyfert 2 galaxy ESO~428-G14 (0714-2914, M4-1). The NLR consists of
many individual, thin strands, which are very closely related to the
radio jet and produce a highly complex yet ordered structure. We find
that the jet is two-sided with a double-helix of emission-line gas
apparently wrapped around the NW side.  To the SE, the jet seems to be
deflected at a ridge of highly excited gas. The strands to the SE may
also wrap around the radio jet, but here complete helices are not
seen. The overall structure is reminiscent of the jet seen in
NGC~4258. Faint symmetric features aligned with the nucleus could
indicate the presence of a highly collimated beam of photons or plasma
from the center.
\end{abstract}

\keywords{galaxies: active --- galaxies: individual (ESO 428-G14) ---
galaxies: jets --- galaxies: nuclei --- galaxies: seyferts --- radio
continuum: galaxies}

\section{Introduction}
The defining features of Seyfert galaxies are their narrow,
high-excitation, nuclear emission-lines, which indicate the presence
of non-stellar, nuclear activity.  The shape and size of the narrow
line region (NLR) are therefore important factors for our
understanding of the processes in Seyfert galaxies in particular and
AGN in general. Observations of the extended narrow line region (ENLR)
have shown that this region is often elongated, and excitation maps
(i.e.~[\ion{O}{3}]/H$\alpha$ ratio maps) sometimes show a sharp,
straight-edged conical structure (e.g.~Pogge 1988). This is well
explained within the unified scheme (see Antonucci 1993), in which
shadowing by an obscuring torus produces an anisotropic radiation
field that excites the ambient medium.

The fact that most NLRs do not show a clear conical structure in total
emission-line maps alone and that these images reveal much fine-scale
structure indicate that other factors, in addition to anisotropy in
ionizing radiation, affect the morphology of the NLR and the clouds
therein.  Neither outflow nor pure rotation seems to provide a complete
description of the kinematics of the NLR clouds (e.g.~in NGC~4151,
Boksenberg et al.~1995), which leaves the important question of what
produces and shapes the NLR unanswered.

Since the early VLA surveys of Seyfert galaxies (Ulvestad \& Wilson
1989 and references therein) it has been known that a substantial
fraction of Seyfert galaxies show radio jets or lobes, and many others
have radio cores that may just be unresolved radio jets. Moreover, in
Seyferts (de Bruyn \& Wilson 1978), radio galaxies (Rawlings et
al.~1989) and quasars (Baum \& Heckman 1989; Falcke, Malkan, Biermann
1995) the emission-line and radio luminosities are
correlated. Further, the radio structure and the (E)NLR are closely
aligned in Seyfert galaxies (Unger et al.~1987, Haniff, Wilson \& Ward
1988). Thus, the physical processes which define the relationship
between radio synchrotron and thermal gases are of considerable
interest.

Hubble Space Telescope (HST) observations of Seyferts (e.g.~Mulchaey et
al.~1994, Bower et al.~1994, Boksenberg et al.~1995, Capetti et
al.~1996) show that the radio emission and the NLR are indeed closely
associated.  In this paper, we present HST/WFPC2 images of the Seyfert
2 galaxy ESO~428-G14 which shed further light on this association
and especially on the interaction between the jet and the NLR.

ESO~428-G14 (0714-2914, MCG -05-18-002) was initially classified as a
planetary nebula (M4-1), but Bergvall et al.~(1986) showed that it is
a typical Seyfert 2 galaxy.  The host galaxy is classified as SA0+. On
high resolution VLA maps, the galaxy shows an apparently one-sided,
bent radio jet (Ulvestad \& Wilson 1989) and Wilson \& Baldwin (1989,
hereinafter WB89) found that the ionized gas, as imaged in
[\ion{O}{3}] and H$\alpha$+[\ion{N}{2}], of ESO~428-G14 is elongated
in the direction of the radio jet. The [\ion{O}{3}] emission-lines
have extended red wings in the NW and blue wings to the SE.  The
heliocentric redshift is $z=0.00544$, and 1\arcsec{} corresponds to 92
parsec for a Hubble constant of $H_0=75\,{\rm km\,s}^{-1}\,{\rm
Mpc}^{-1}$ using the Galactic Standard of Rest velocity from de
Vaucouleurs et al.~(1992).

\section{Observations and Data Reduction}
\subsection{Observations}
ESO~428-G14 was observed with the Wide Field and Planetary Camera 2
(WFPC2) on board the HST.  The filters were chosen such that they
included the redshifted H$\alpha$ ($\lambda$6563\AA{}) and
[\ion{O}{3}] ($\lambda5007$\AA) emission-lines and their adjacent
continua, all exposures being split into two integrations, to allow cosmic ray
rejection.

The H$\alpha$ emission-line of ESO~428-G14 is redshifted into the
[\ion{N}{2}] filter (F658N) of the WFPC2 and so we were able to use
the PC chip which has a pixel scale of 0\farcs0455/pixel. Using the
spectrum from Bergvall et al.~(1986) of ESO~428-G14 and the bandpass
of the F658N filter, one finds that [\ion{N}{2}] (mostly the
$\lambda6548$\AA{} line) contributes $\simeq33\%$ of the emission-line
flux in this image. For simplicity we will refer to our F658N images
as ``H$\alpha$ images'' despite the [\ion{N}{2}] contribution and no
correction was applied for the [\ion{N}{2}] contamination, as the
spatial variation of the [\ion{N}{2}]/H$\alpha$ ratio is unknown. For
the continuum a red broadband filter (F814W) was chosen, in which
little contamination from emission-lines is expected.  The
observations were performed on 1995 April 17, and had total exposure
times of 800 s and 200 s in the F658N and F814W filters respectively.

For the [\ion{O}{3}] image, the Linear Ramp Filters (LRF) and the WFC chips
with a pixel scale of 0.1\arcsec{}/pixel were used.  The LRFs have
position-dependent peak-transmission wavelengths and FWHM bandpasses
of $\sim1.3\%$, and so they are ideally suited for redshifted
emission-line objects which are less than 14\arcsec{} across.  The
filter used for the redshifted [\ion{O}{3}] image was FR533N with the
galaxy centered at position (675,227) on the WF4 chip which corresponds
to a central wavelength of $\lambda5033$\AA{}. The off-band continuum
image was taken on the WF3 chip centered at position (221,565),
which corresponds to a central wavelength of $\lambda5284$\AA{}. Those
observations were performed on 1996 January 14 with total exposure
times of 280 s each.

\subsection{Data Reduction}

All images were processed through the WFPC2 pipeline at the Space
Telescope Science Institute. Flatfields are not yet available for the
LRFs, so we used a flatfield taken in a nearby narrow-band filter
(F502N). Each pair of images was combined using the {\sc iraf} {\tt
crrej} task to remove cosmic rays, and photometric calibration was
performed using the WFPC2 exposure time calculator.  All images were
rotated to the cardinal orientation.

Combining the LRF images taken at different positions on the WFPC
requires correction for geometric distortions, and this was achieved
using the {\sc iraf} {\tt wmosaic} task. Each pair of line and
continuum images was aligned using the internal astrometry of the HST,
i.e.~using the coordinate information in the image headers. Since each
of the on- and off-band pairs was taken consecutively using the same
set of guide stars, the relative astrometry should be excellent. The
positions of three stars visible on the red on- and off-band images
were found to agree to better than 20 mas.

It was then necessary to align the PC (red) and LRF (green) images.
This was done by aligning the continuum peaks, which are well defined
in our images, and the corresponding on-band images were shifted
accordingly. As we did not have any stars common to both images, we
checked the registration by comparing the very similar morphology in
the two on-band images. The good agreement indicates that the position
of the optical continuum position (see crosses in Fig.~1a \& b) in both
images relative to each other is not severely altered by reddening. To
obtain an excitation map, the [\ion{O}{3}] image was interpolated to
match the sampling of the H$\alpha$ image. Then both images were
smoothed with a beam of 0\farcs1 and divided by each other. The
resulting map was similar to that obtained by reducing the PC image to
the resolution of the WF image prior to the division.

\section{Results}
\subsection{HST images}
Our final images are shown in Fig.~1 (Plate X) which reveal that the
narrow line region of ESO~428-G14 is very asymmetric, highly elongated
and has very prominent, bright strands along its major axis. As
commonly found for Seyfert 2 galaxies observed with HST (e.g.~Wilson
et al.~1993, Bower et al.~1994), the nucleus of the galaxy (taken here
to be the peak of the red continuum light) is not particularly bright
in emission-lines. One of the most interesting features of the
emission-line maps (Fig.~1a \& b) is the morphology of the ionized gas
seen to the NW of the nucleus, which resembles a ``figure of eight'' or two
overlapping, S-shaped emission-line strands.  The SE part of the NLR
is markedly different: the NLR is much narrower close to the nucleus,
extends at least twice as far from the nucleus as the NW part, and
towards the end splits into several strands of emission-line gas which
turn towards the north. Despite these differences in detail, the
emission-line gas on both sides of the nucleus is dominated by
partially resolved strands which are only 10-20 pc wide, but 100-300
pc long. Because they are so narrow, the strands are best seen on the
H$\alpha$ image (Fig.~1a), which has higher spatial resolution than
the [\ion{O}{3}] image.

Besides the strands we find a few fainter structures in our
emission-line images that are of potential interest. There is a faint
arc of emission-line gas some 1\arcsec{} (100 pc) south of the east end
of the emission-line strands, best seen in the [\ion{O}{3}] image. On
the H$\alpha$ image we notice two weak blobs 5\arcsec{} to the SE and
3\arcsec{} to the SSE of the nucleus.  On the other side of the NLR, a
faint, linear emission-line feature extends from the ``figure of eight''
to the NW. It is $\sim1.1\arcsec{}$ long and points directly towards
the nucleus. At the same distance (2.2\arcsec{} from the nucleus) in
the opposite direction, the emission from the lower part of the
emission-line arc is also slightly enhanced.

Interestingly, these linear features to the NW and SE, the nucleus
itself, and the 5\arcsec{} SE H$\alpha$ blob align to within a few
degrees. The odds for a chance alignment are minimal given the fact
that we see only two isolated H$\alpha$ blobs within the central
10\arcsec{}x10\arcsec{} and the question arises whether the blob
5\arcsec\ SE of the nucleus is excited or has been expelled by the
nuclear source. We therefore smoothed the [\ion{O}{3}] and H$\alpha$
images with a 0.2\arcsec{} beam to increase the signal to noise ratio,
and determined the [\ion{O}{3}]/H$\alpha$ ratio for the blob and the
linear features. While the linear features have
[\ion{O}{3}]/H$\alpha\ga2$, consistent with being excited by an AGN
spectrum, the SE blob has [\ion{O}{3}]/H$\alpha\sim1/3$.

To learn more about the excitation, we have divided the [\ion{O}{3}]
image by the H$\alpha$ image at the resolution of the WF pixels. In
Fig.~1c, regions of darker shades indicate higher
[\ion{O}{3}]/H$\alpha$ ratios and hence regions of higher
excitation. The excitation structure to the SE of the nucleus and
within 2\arcsec{} of it bears some resemblance to the ionization cones
seen in other Seyfert 2 galaxies. If one interprets this structure as
being due to a cone of exciting radiation, the opening angle would
be around $45-50^\circ$. The structure on the other side of the
nucleus is consistent with the presence of a similar, but shorter cone.
The average [\ion{O}{3}]/H$\alpha$ ratio along the central axis of the
cone is around 2. There is, however, a noticeable trend for the
excitation to decrease towards the NW. This could be understood as a
result of increased reddening in this direction.

Towards the SE end of the nebulosity, in the region where the radio
jet and the last strand bend towards the north, the excitation in the
gas gas is relatively high. The average [\ion{O}{3}]/H$\alpha$ here is
around 3.2. The faint arc to the south of the east end of the strands
also has higher than average excitation, but here the data are noisy.
The highest excitation found in this map -- close to the nucleus -- is
around 3.5.

Finally, we present the red broad-band image of the galaxy in Fig.~1d.
Within the inner 100 pc, a spiral feature extends from the nucleus to
the SE and then turns towards the north. The continuum nucleus is clearly
elongated on a scale of a few tenths of an arcsecond along the
direction of the NLR, and further out the contours are disturbed at
the position of the bright emission-line strands. The galaxy extends
further than the section shown in Fig.~1d, but even on the full field
of view, we do not see any large scale spiral arms.

\subsection{Comparison with radio map}
The absolute astrometric uncertainties for HST positions with respect
to the radio reference frame are of the order 1\arcsec{} and hence
cannot be used to align the HST and the VLA maps of
ESO~428+G14. However, we smoothed the HST image to the resolution of
the radio map (0.3$\times$0.6\arcsec{}) and found that the smoothed
emission line image is very similar to the radio map. If we then make
the reasonable assumption that the radio and optical structures should
align, we can shift the optical map until the structures
coincide. Even though the radio and optical emission are sometimes
anti-correlated in detail (e.g.~Mulchaey et al.~1994) the similarity
of the radio continuum and optical emission-line structures of
ESO~428-G14 suggests that this is not a major problem here.

However, within an accuracy of 0.2\arcsec{}, the pixel-to-pixel
registration for the two maps becomes rather arbitrary. To make a
quantifiable registration, we have arbitrarily aligned the second
brightest radio blob (some 0\farcs75 SE of the brightest radio
feature) with the optical continuum peak. Fig.~2a shows the contours
of the radio map of Ulvestad \& Wilson (1989) overlaid on the shifted
H$\alpha$ image with this alignment. The (B1950) coordinate grids on
Figs.~1 were also taken from the radio map. With this final alignment,
the shift between the HST and the VLA coordinate grid are 1.0\arcsec{}
in RA and 0.4\arcsec in DEC, and hence within the typical astrometric
uncertainties of the HST.

Provided our registration is correct, the first result of this
comparison is that the brightest peak of the radio emission is
associated with the NW part of the NLR (the ``figure of eight'') and
is displaced by $\sim$0.75\arcsec{} from the optical continuum peak.
We therefore conclude that the radio jet is indeed not one-sided, but
rather two-sided and asymmetric, as speculated by WB89.  Like the
emission-line structure, the NW side of the radio jet is very short
and terminates in a bright hotspot, while there is no well defined
terminus to the SE, where the radio jet and emission-line strands bend
towards the north.  However, the somewhat higher excitation seen in
the [\ion{O}{3}]/H$\alpha$ map (Fig.~2b) along the the bent of the
radio jet in the SE may indicate some increased jet/ISM interaction.

The only structure which does not have a correspondence in the radio
map is the emission-line arc to the south of the SE tip of the
NLR. Because of the lower resolution of the radio map and the
uncertainty of registration, we also cannot tell whether the
emission-line strands themselves are associated with radio emitting
plasma.

\section{Discussion}
In general ESO~428-G14 fits quite nicely into the basic picture we
have for Seyfert 2 galaxies within the unified scheme. The ionization
structure as seen in the excitation map is highly elongated and
consistent with a faint bi-cone, even though the structure is less
convincing than in the best cases of ``ionization cones''
(e.g. Tadhunter \& Tsvetanov (1989).  We also confirm the results of
WB89 that the emission-line gas is elongated in the same direction as
the radio outflow, which seems to be a general feature of Seyfert
galaxies.

The most remarkable result for ESO~428-G14, however, is the shape of
the NLR and its close correspondence to the radio jet, which allowed
us to align radio and optical maps fairly accurately.  The emission
from the NLR of ESO~428-G14 is dominated by well ordered strands of
emission-line gas which are some 10 parsec thick, but can be more than
100 pc long. In the NW part of the jet, we find two strands organized
in a ``figure of eight'' shape that is very suggestive of a
double-helix.  This may indicate that the emission-line strands are
produced at the surface of the radio jet, possibly by an interaction
of the jet with the ISM.  Helical structures associated with radio
jets seem to be quite common: some VLBI components of quasar jets seem
to move on helical trajectories (see e.g.~Steffen et al.~1995), and
Cecil et al. (1992) demonstrated that the H$\alpha$ jet in NGC~4258
consists of helically-twisted strands in a triple helix.

The comparison with NGC~4258 is even more striking if one considers
the region to the SE, where the jet tapers off. The strands in
NGC~4258 become detached from each other, leave the helical structure
and end up pointing in different directions. This is similar to the SE
part of the ESO~428-G14 jet, where the relatively narrow emission-line
structure splits into various strongly bent fingers. Moreover, in both
galaxies, the untwisting of the strands happens at the end of the jet
and is associated with strong jet bending.  This similarity may
suggest that even if we cannot resolve the well collimated structure
just to the SE of the nucleus of ESO~428-G14, there may well be a
tightly twisted string of helical strands, as in NGC~4258.  On the
other hand, it is striking that the helices on opposite sides of the
nucleus of ESO~428-G14 should have so very different pitch angles.  It
may therefore be that we see only one strand of a wide helix in the
southern jet, which then splits into multiple strings.

What processes can be responsible for these helical strands in
ESO~428-G14? The only plausible place to produce such strands is the
boundary layer between the radio jet and the ISM. In NGC~4258, it has
been argued that the jet creates a low density tunnel through the ISM
(Martin et al.~1989). A similar process may occur in ESO~428-G14.
Interactions between the jet and the denser ISM could then induce
fluid instabilities and entrainment in the jet --- the pros and cons
of these models are discussed extensively in Cecil et al.~(1992).

Another question that immediately comes to mind when looking at
Fig.~1a is whether the optical emission-line strands could be
associated with strong magnetic field lines.  Those field lines would
have to be frozen into the plasma on the surface of the jet which is
ionized by the central source (the fact that the strands do not show
up in the excitation map suggests that there is little or no local
ionization). Such prominent, large scale magnetic field lines do indeed exist, as
seen for example in the Galactic Center non-thermal filaments
(Yusef-Zadeh \& Morris 1987) or in other exotic radio sources
(e.g. Gray 1996).

In any case it appears from our images that the morphology of the NLR
is largely governed by the bipolar outflow. WB89 have found that the
emission-lines in ESO~428-G14 have extended red wings to the NW of the
nucleus, and extended blue wings to the SE, with speeds up to 1400 km
s$^{-1}$.  This suggests, that the NW jet is moving away from us and
the SE jet towards us. Such an orientation would be consistent with
the apparently larger reddening to the NW which we inferred from the
lower [\ion{O}{3}]/H$\alpha$ ratio in the NW (Sec.~3.1).

Besides the strands, we find other signs of the interaction between
the jet and the ISM. As mentioned before, along the large scale bend
of the SE jet we find a ridge of enhanced excitation. An association
between jet terminus and enhanced [\ion{O}{3}] emission has also been
found in many radio galaxies (McCarthy, Spinrad, \& van Breugel
1995). The faint arc of emission-line gas we see beyond this ridge is
reminiscent of the spectacular arcs found in the Seyfert galaxy
Mrk~573 (Capetti et al.~1996) where they were interpreted as
shockfronts. However, in ESO~428-G14 the jet does not even seem to
reach this arc and deeper radio maps are need to investigate this
phenomenon in more detail.

Finally, we consider the two faint, linear features which are
symmetrically placed in opposite directions 2.2\arcsec{} from the
nucleus --- the NW feature is well visible in Fig.~1a, while the SE
feature is on top of the faint arc and less apparent. The presence of
these features is suggestive of a highly collimated beam of ionizing
photons escaping from the center. In NGC~3516 a similar feature
pointing towards the nucleus was seen by Miyaji, Perez-Fournon, \&
Wilson (1992) and cautiously interpreted as ambient gas illuminated by
UV radiation from a relativistic jet. The low-excitation blob with
cometary-like tail 5\arcsec{} SE of the nucleus is aligned in the same
directions, but in view of its low excitation, this blob might be just
an \ion{H}{2}-region.

Despite the many open questions, ESO~428-G14 may contribute
significantly to our understanding of the NLR of AGN and the jet/ISM
interaction. The galaxy shows the main features seen in various
Seyfert and radio galaxies, i.e.~a possible ionization cone and a
close association between radio synchrotron and optical emission-line
gases. It also demonstrates that the structure of the NLR cannot be
understood without considering the influence of the jet on the
ISM. Higher resolution radio maps are needed to determine whether the
strands of ionized gas are also sources of enhanced synchrotron
emission.

\acknowledgements This research was supported by NASA under grants
NAGW-3268 and NAG8-1027. We thank the STScI staff -- especially
J. Biretta, M. McMaster, and K. Rudloff -- for their support in the
LRF calibrations, and an anonymous referee for helpful remarks.

\clearpage
\figcaption[]{}(a) Continuum subtracted H$\alpha$ image of the narrow
line region of ESO~428-G14 taken with the Planetary Camera
(0\farcs0544 resolution). The flux in the central
4\arcsec$\times$4\arcsec\ is $8.2\cdot10^{-13}$ erg s$^{-1}$ cm$^{-2}$
and the intensity scale here and in the following is proportional to
the square root of the brightness. The continuum peak of the galaxy is
marked by a cross, and the (B1950) coordinates are from the 6 cm radio
map (see Fig.~2).  (b) Continuum subtracted [\ion{O}{3}] image of the
narrow line region of ESO~428-G14 taken with the Wide Field Camera
(0\farcs1 resolution).  The flux in the central
4\arcsec$\times$4\arcsec\ is $1.5\cdot10^{-12}$ erg s$^{-1}$ cm$^{-2}$.
(c) Excitation map of ESO~428-G14, obtained by dividing
[\ion{O}{3}]/H$\alpha$. Dark shades represent regions of high
excitation, the grayscale ranges from 1.2 to 4 for the
[\ion{O}{3}]/H$\alpha$ ratio.  (d) Red continuum filter (F814W) image
of the host galaxy overlaid with logarithmic isophotes (0\farcs0544
resolution).

\figcaption[]{}H$\alpha$ and excitation map (both with linear
greyscale) of ESO~428-G14 as in Fig.~1 overlaid with the contours of
the 6cm map from Ulvestad \& Wilson (1989), which has a beam of
$0.3''\times0.6''$. The (B1950) coordinate grid is also taken from
this map. The uncertainty in the registration between VLA and HST map
are $\sim2$\arcsec.
\clearpage

\onecolumn
\begin{figure*}
\plotone{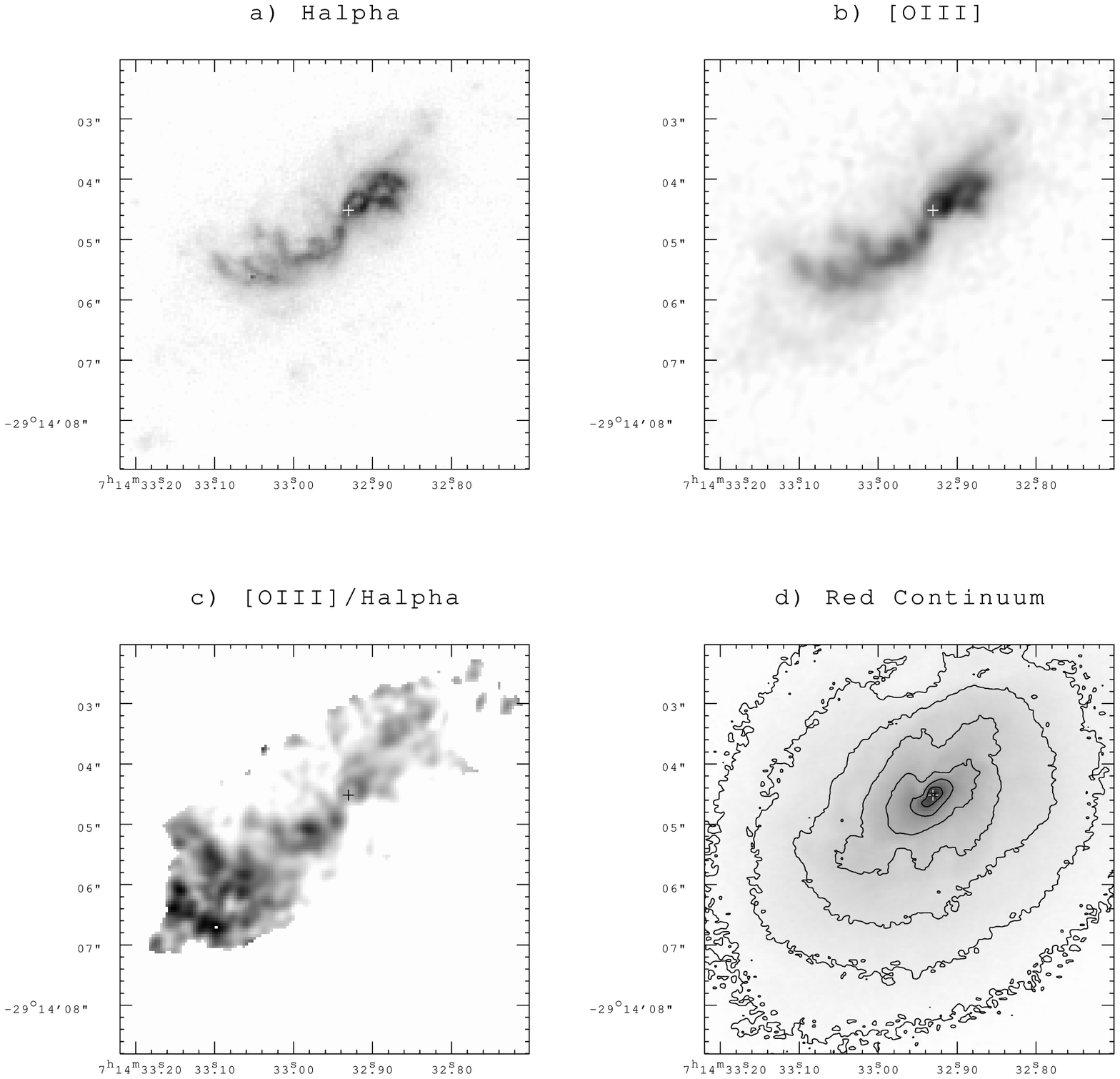}

\centerline{Figure 1 (Plate N)}
\end{figure*}

\clearpage
\begin{figure*}
\plotone{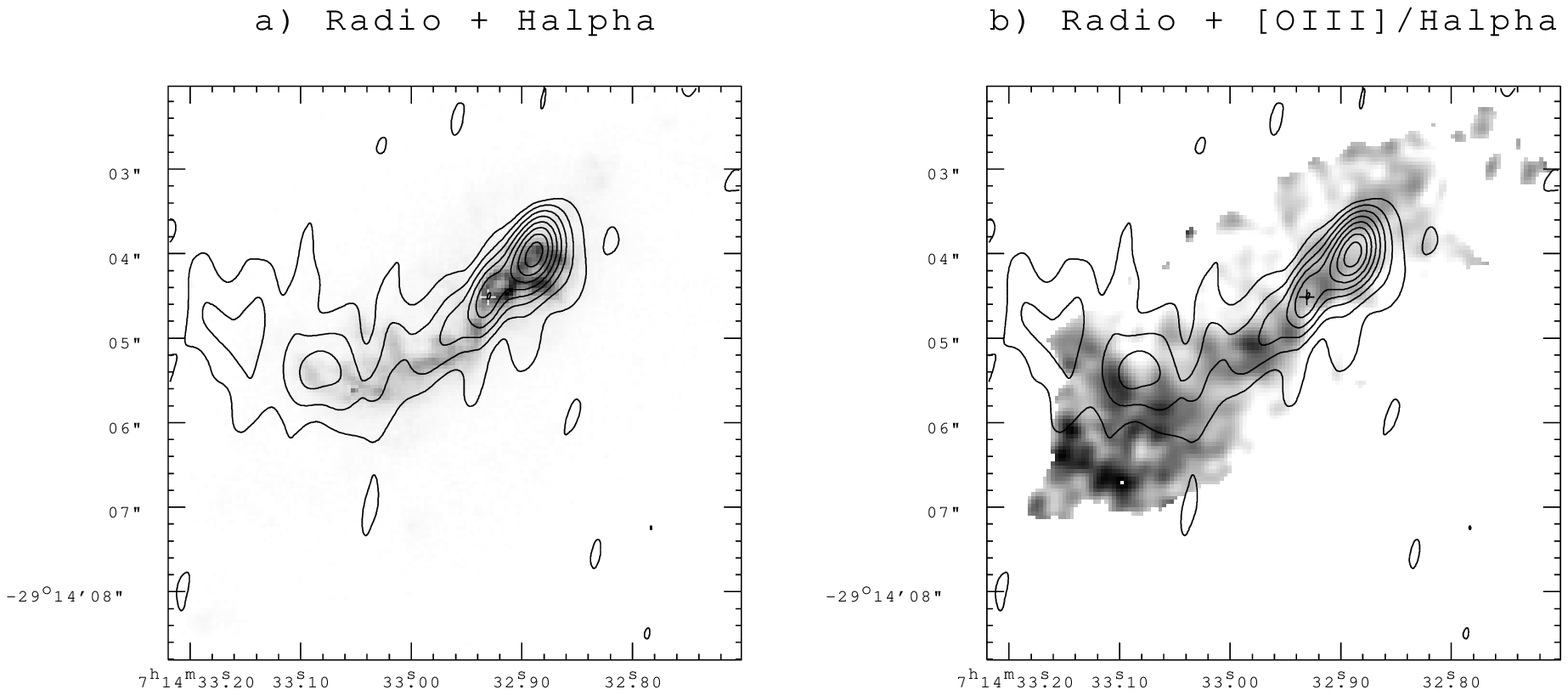}

\centerline{Figure 2}
\end{figure*}

\end{document}